\documentclass[aps,prb,twocolumn,showpacs,preprintnumbers,amsmath,amssymb,groupedaddress,noshowkeys]{revtex4}
\usepackage{txfonts}
\usepackage[dvipdfm]{graphicx}
\usepackage{bm}

\def\rnum#1{\expandafter{\romannumeral #1}} 
\def\Rnum#1{\uppercase\expandafter{\romannumeral #1}}

\newcommand{\cre}[2]{{#1}_{#2}^{\dagger}}
\newcommand{\ani}[2]{{#1}_{#2}^{\ }}
\newcommand{\p}{\prime}
\newcommand{\da}{\dagger}

\newcommand{\ef}{E_\text{F}}
\newcommand{\kf}{k_\text{F}}

\newfont{\bg}{cmr10 scaled\magstep4}

\newcommand{\bigzerou}{\smash{\lower1.8ex\hbox{\bg 0}}}

\begin{document}

\title{Enhanced spin Hall effect by tuning antidot potential:
Proposal for a spin filter}
\author{Tomohiro Yokoyama}
\email[E-mail me at: ]{tyokoyam@kobrmc34.rk.phys.keio.ac.jp}

\author{Mikio Eto}
\affiliation{Faculty of Science and Technology, Keio University,
3-14-1 Hiyoshi, Kohoku-ku, Yokohama 223-8522, Japan}
\date{\today}       

\begin{abstract}
We propose an efficient spin filter including an antidot
fabricated on semiconductor heterostructures with strong
spin-orbit interaction. The antidot creates a tunable
potential on two-dimensional electron gas in the
heterostructures, which may be attractive as well as repulsive.
Our idea is based on the enhancement of extrinsic spin Hall
effect by resonant scattering when the attractive
potential is properly tuned. Numerical studies for three-
and four-terminal devices indicate that the efficiency of
the spin filter can be more than 50\% by tuning the
potential to the resonant condition.
\end{abstract}
%\keywords{spin-orbit interaction; spin Hall effect; antidot;
%resonant scattering}
%\preprint
\maketitle

\section{INTRODUCTION}

The injection and manipulation of electron spins in
semiconductors are important issues for spin-based electronics,
``spintronics.''\cite{Zutic} The spin-orbit (SO) interaction
plays an important role in
the manipulation of the spins.
For conduction electrons in direct-gap semiconductors,
the SO interaction is expressed in the same form as that
in vacuum,
that is,
\begin{equation}
H_{\rm SO} =\frac{\lambda}{\hbar} \bm{\sigma} \cdot
\left[\bm{p} \times \bm{\nabla} V(\bm{r}) \right]
\label{eq:SOorg}
\end{equation}
where $V(\bm{r})$ is an external potential and $\bm{\sigma}$
indicates the electron spin $\bm{s}=\bm{\sigma}/2$.
The coupling constant $\lambda$ is significantly enhanced by
the band effect, particularly in narrow-gap semiconductors such
as InAs,~\cite{Winkler} compared with that in vacuum,
$|\lambda|=\hbar^2/(4 m_0^2 c^2)$ with $m_0$ as the
electron mass and $c$ as the velocity of light.

In two-dimensional electron gas (2DEG) in semiconductor
heterostructures, an electric field perpendicular to the
2DEG results in the Rashba SO interaction.~\cite{Rashba} 
For the electric field ${\cal E}$ in the $z$ direction,
the substitution of $V(\bm{r})=e {\cal E} z$ into Eq.\
(\ref{eq:SOorg}) yields
\begin{equation}
H_{\rm SO} =\frac{\alpha}{\hbar} (p_y\sigma_x-p_x\sigma_y),
\label{eq:Rashba}
\end{equation}
with $\alpha=e {\cal E} \lambda$. Large values of $\alpha$
have been reported in experiments.\cite{Nitta,Grundler,Yamada}
In the spin transistor proposed by Datta and
Das,~\cite{spintransistor} electron spins are injected into
the semiconductor heterostructures from a ferromagnet,
and manipulated by tuning the strength of Rashba SO
interaction by adjusting the electric field ${\cal E}$.
The spins are detected by another ferromagnet.
It is well-known, however, that the efficiency of spin
injection from a ferromagnetic metal to semiconductors is
very poor, less than 0.1\%, due to the conductivity
mismatch.~\cite{mismatch}
To overcome this difficulty,
the SO interaction may be useful for
the efficient spin injection, besides the spin manipulation,
in the spin transistor.
Several spin filters have been
proposed utilizing the SO interaction, e.g.,
three- or four-terminal devices based on the spin Hall
effect (SHE),~\cite{Bulgakov,Kiselev,Kiselev2,Pareek}
a triple-barrier tunnel diode,~\cite{3diode}
%%%%%%%%%%
a quantum point contact,~\cite{Eto05,Silvestrov}
a three-terminal device for the Stern-Gerlach experiment
using a nonuniform SO interaction,~\cite{3termSG}
and an open quantum dot.~\cite{Krich}
Yamamoto and Kramer proposed a three-terminal spin filter
with an antidot, using a SHE caused by the scattering
of electrons at a repulsive potential created by the
antidot.~\cite{Yamamoto}

The SHE is one of the phenomena to create a spin current due
to the SO interaction. There are two
types of SHE.
One is an intrinsic SHE which creates a dissipationless spin
current in the perfect crystal.~\cite{Murakami,Sinova}
Murakami \textit{et al}., for example, proposed that
the drift motion of holes in SO-split valence bands induces
an intrinsic SHE.~\cite{Murakami}
The SHE of the hole system has been observed experimentally
by Wunderlich \textit{et al}.,
using a $p$-$n$ junction light-emitting diode.~\cite{Wunderlich}
The other type is an extrinsic SHE caused by the spin-dependent
scattering of electrons by
impurities.~\cite{Dyakonov,Hirsch,Zhang,Engel}
For a centrally symmetric potential around an impurity, $V(r)$
($r=\sqrt{x^2+y^2+z^2}$), Eq.\ (\ref{eq:SOorg}) is rewritten as
\begin{equation}
H_{\rm SO} =-\lambda\frac{2}{r} \frac{dV}{dr}
\bm{l} \cdot \bm{s},
\label{eq:SO3D}
\end{equation}
where $\bm{l}=(\bm{r} \times \bm{p})/\hbar$ is the angular momentum.
This results in the skew scattering: accompanied by the
scattering from direction $\bm{n}$ to $\bm{n'}$, the spin is
polarized in $(\bm{n} \times \bm{n'})/
|\bm{n} \times \bm{n'}|$.\cite{Mott,Landau}
In an optical experiment on the Kerr rotation, Kato {\it et al}.\
observed a spin accumulation at sample edges along
the electric current in $n$-type GaAs,\cite{Kato}
which is ascribable to the extrinsic SHE.
The experimental result has been quantitatively explained
by a semi-classical theory considering the skew scattering
and ``side jump'' effects.~\cite{Engel}

In our previous paper,~\cite{Eto} we have quantum-mechanically
formulated the extrinsic SHE for 2DEG in semiconductor
heterostructures. We have examined the SHE due to the scattering
by an artificial potential created by an antidot,
scanning tunnel microscope (STM) tip, etc.
An antidot is a small metallic electrode fabricated above
2DEG, which creates an electrically tunable potential on 2DEG.
The potential is usually repulsive, but it could be attractive
when a positive voltage is applied to the antidot.
We have found that the
SHE is significantly enhanced by resonant scattering
when the attractive potential is properly tuned.

We have stressed that the extrinsic SHE is easier to understand
in 2DEG than in three-dimensional case. Let us consider
an axially symmetric potential $V(r)$ ($r=\sqrt{x^2+y^2}$)
created by an antidot on conduction electrons
in the $xy$ plane.
The SO interaction in Eq.\ (\ref{eq:SOorg}) becomes
\begin{equation}
H_\text{SO} = -\lambda \frac{2}{r} \frac{dV}{dr} l_z s_z
\equiv V_1(r) l_z s_z,
\label{eq:SO2D}
\end{equation}
where $l_z$ and $s_z$ are the $z$ component of angular momentum
and spin operators, respectively.
$V_1(r)=-(2 \lambda/r)V'(r)$, which has the
same sign as $V(r)$ if $|V(r)|$ is a monotonically decreasing
function of $r$ and $\lambda>0$. Assuming that $V(r)$ is
smooth on the scale of the lattice constant, we adopt the
effective mass equation
\begin{equation}
\left[ -\frac{\hbar^2}{2m^*} \Delta + V(r) + V_1(r)l_z s_z \right]
\psi({\bm r}) =E \psi({\bm r}),
\label{eq:Schroedinger}
\end{equation}
for an envelope function $\psi({\bm r})$ with $m^*$ as the
effective mass. In Eq.\ (\ref{eq:Schroedinger}),
$l_z$ and $s_z$ are conserved
in contrast to the three-dimensional case with
Eq.\ (\ref{eq:SO3D}). For $s_z=\pm 1/2$, an electron experiences
the potential of $V(r) \pm V_1(r) l_z/2$. As a consequence, the
scattering of components of $l_z>0$ ($l_z<0$) is enhanced
(suppressed) by the SO interaction for $s_z=1/2$ when
$V_1(r)$ has the same sign as $V(r)$.
The effect is opposite to that for $s_z=-1/2$. This is the origin
of the extrinsic SHE.
We have formulated the SHE in terms of phase shifts in the
partial wave expansion for 2DEG and shown that the SHE
is largely enhanced by resonant scattering.~\cite{Eto}
These results are summarized in Appendix A.

\begin{figure}
\includegraphics[width=8.5cm]{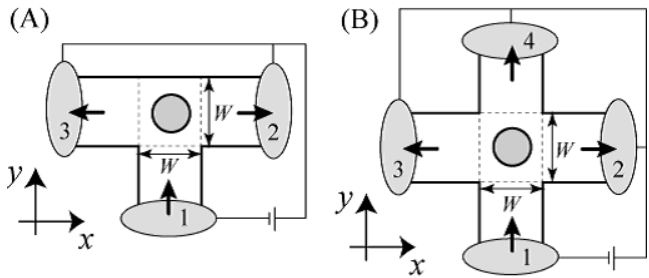}
\caption{Our model for (A) three- and (B) four-terminal devices
for the spin filter. They are fabricated on two-dimensional
electron gas in the $xy$ plane. Both the devices include an antidot
at the center of junction, which is a square area surrounded
by broken lines. Three or four ideal leads connect the junction
to reservoirs. Reservoir 1 is a source from which unpolarized
electrons are injected into the junction.
The voltages are equal in the drains; reservoirs 2 and 3 in model
(A) and reservoirs 2, 3, and 4 in model (B).}
\label{fig:System}
\end{figure}

In the present paper, we consider three- and four-terminal
devices including an antidot, as shown in
Fig.\ \ref{fig:System}, and examine the enhancement of
the SHE. We evaluate an efficiency of the
spin-filtering effect by resonant scattering in the case
of attractive potential. Although our three-terminal device
is very similar to the spin filter proposed by
Yamamoto and Kramer,~\cite{Yamamoto}
they have only studied the case of repulsive potential.
We show that our device can be a spin filter with an
efficiency of more than 50\% by tuning the potential to
the resonant condition.

%%%figure
%%%%%%%%%%%%%

We numerically solve the effective mass equation
[Eq.\ (\ref{eq:Schroedinger})] with an appropriate boundary condition
for our devices.
A confining potential for the leads (quantum wires) could induce
the SO interaction, following Eq.\
(\ref{eq:SOorg}).~\cite{Hattori,Bellucci,Jiang,Xing}
However, its effect on the electrons is much smaller than
the SO interaction induced by the antidot potential because the
amplitude of the wavefunction is small around the edges of the
leads. Therefore, we consider the antidot-induced SO interaction
only.
We also assume that the antidot potential $V(\bm{r})$ is
independent of $z$. Otherwise, it would create the
Rashba-type SO interaction, Eq.\ (\ref{eq:Rashba}) with
$\alpha=\lambda(\partial V/\partial z)$, in addition to
Eq.\ (\ref{eq:SO2D}).
The Dresselhaus SO interaction is also disregarded, which is
induced by the inversion asymmetry of the crystal.~\cite{Dresselhaus}
These effects will be discussed in the final section.

%%%%%%%%%%%%%%
The electron-electron interaction is not taken into account.
The Coulomb blockade is not relevant to
the case of antidot, in contrast to that of conventional
quantum dot which is connected to the leads via tunnel
barriers.~\cite{Kouwenhoven} In our model, therefore, the influence of
the electron-electron interaction is only quantitative and
could be considered by the mean-field level, as discussed in
the final section. Note that there have been
several researches for the spin-current generation based on
the electron-electron interaction in the absence of SO
interaction, e.g., using single or
double quantum dots~\cite{Aono,Feinberg,Pustilnik1}
and quantum wires.~\cite{Sharma,Citro,Pustilnik2,Braunecker,Abanin}

The organization of the present paper is as follows.
In Sec.\ II, we describe our model for three- and four-terminal
devices and calculation method. The calculation of spin-dependent
conductance in multi-terminal devices is formulated using the
Green's function in the
tight-binding model.~\cite{Datta,Ando,Ando2,Yamamoto2}
In Sec.\ III, we present numerical results of
the conductance and spin-filtering effect when the strength of
antidot potential is tuned. We also investigate the density
of states (DOS) in the junction area of the devices to
illustrate virtual bound states. The existence of the
virtual bound states at the Fermi level is a strong evidence
that a resonant scattering takes place when the
spin-filtering effect is remarkably enhanced.
In addition, we perform a channel analysis of the
spin-dependent conductance to closely examine the resonance.
The final section (Sec.\ IV) is devoted to the conclusions
and discussion.

\section{MODEL AND CALCULATION METHOD}

In this section, we explain our model and calculation method.
We numerically solve the effective mass equation in the
tight-binding model for the devices. In the presence of the SO
interaction in Eq.\ (\ref{eq:SO2D}), the $z$ component of
electron spin $s_z$ is conserved although $l_z$ is not a good
quantum number owing to the lack of rotational symmetry in
our devices. Hence we can solve the equation for
$s_z=\pm 1/2$ separately.

\subsection{Model}

We consider three- and four-terminal devices with an antidot,
fabricated on semiconductor heterostructures, as shown
in Fig.\ \ref{fig:System}.
Three or four leads (quantum wires) of width $W$ are joined
to one another at a junction, which is a square area
surrounded by broken lines in the figure.
The leads are formed by hard-wall potential and
connected to the reservoirs. Reservoir 1 is a source from which
unpolarized electrons are injected into the junction.
The electrons are outgoing to the drains; reservoirs 2 and 3
(2, 3, and 4) in the three-terminal (four-terminal) device.
The voltages are equal in all the drains.

An antidot creates an axially symmetric potential $V(r)$, where $r$
is the distance from the center of the junction. It is assumed to
be attractive and given by a smooth potential well,
\begin{equation}
V(r)= \begin{cases}
       V_0 & \text{($r-R_0 <-\frac{\Delta R_0}{2}$)} \\
       \frac{V_0}{2} \left\{ 1-\sin \left( \pi
       \frac{r-R_0}{\Delta R_0} \right) \right\}
       & \text{($|r-R_0 |\le \frac{\Delta R_0}{2}$)} \\
       0 & \text{($r-R_0 >\frac{\Delta R_0}{2}$)}
       \end{cases}
\label{eq:potential}
\end{equation}
with $V_0<0$.
The radius of the potential well is $R_0 =W/4$, and we choose
$\Delta R_0 =0.7 R_0$. The gradient of $V(r)$ gives rise to the
SO interaction in Eq.\ (\ref{eq:SO2D}).

For the numerical study, we discretize the two-dimensional space
with lattice constant $a$ (tight-binding model with a square
lattice). The width of the leads is $W=(N+1)a$ with $N=29$;
the wavefunction becomes zero at the zeroth and ($N+1$)th sites in
the transverse direction of the leads. The Hamiltonian is
\begin{equation}
 \begin{split}
 H &=t \sum_{i,j,\sigma} (4+\tilde{V}_{i,j}^{\ })
       \cre{c}{i,j;\sigma} \ani{c}{i,j;\sigma} \\
 & -t \sum_{i,j,\sigma} (T_{i,j ; i+1,j; \sigma}^{\ }
 \cre{c}{i,j;\sigma} \ani{c}{i+1,j;\sigma} \\
 & + T_{i,j ; i,j+1 ;\sigma}^{\ } \cre{c}{i,j,\sigma}
 \ani{c}{i,j+1,\sigma}  + \text{h.\ c.}),
 \end{split}
 \label{eq:AndoH}
\end{equation}
where $\cre{c}{i,j;\sigma}$ and $\ani{c}{i,j;\sigma}$ are
creation and annihilation operators of an electron at site 
$(i,j)$ with spin $\sigma$.
Here, $t=\hbar^2 /2m^{*} a^2$, where $m^{*}$ is the effective
mass of electrons.
$\tilde{V}_{i,j}$ is the potential energy at site $(i,j)$,
in units of $t$. The transfer term in the $x$ direction is given by
\begin{equation}
T_{i,j; i+1,j; \pm} = 1\pm i \tilde{\lambda}
   (\tilde{V}_{i+1/2,j+1} -\tilde{V}_{i+1/2,j-1}),
\label{eq:xhopping}
\end{equation}
whereas that in the $y$ direction is
\begin{equation}
T _{i,j ; i,j+1; \pm} =1\mp i \tilde{\lambda}
   (\tilde{V}_{i+1,j+1/2} -\tilde{V}_{i-1,j+1/2}),
\label{eq:yhopping}
\end{equation}
where $\tilde{\lambda} =\lambda /(4 a^2)$ is the
dimensionless coupling constant of the SO interaction.
$\tilde{V}_{i+1/2,j}$ is
the average of the potential energy at sites
$(i,j)$ and $(i+1,j)$, and $\tilde{V}_{i,j+1/2}$ is
that at sites $(i,j)$ and $(i,j+1)$.

The SO interaction is absent in the leads.
The wavefunction of conduction channel $\mu$ in the leads
is written as
\begin{eqnarray}
\psi_\mu (i^\p , j^\p ) & = & 
\exp(ik_{\mu}a j^\p) u_{\mu} (i^\p),
\\
u_{\mu} (i^\p) & = &
\sqrt{\frac{2}{N+1}} \sin \left( \frac{\pi \mu i^\p}{N+1}\right),
\label{eq:Wavefunc}
\end{eqnarray}
where $i^\p$ and $j^\p$ denote the site numbers in
the transverse and longitudinal directions of the leads,
respectively.
The wavenumber $k_{\mu}$ is determined from the condition that
the dispersion relation
\begin{equation}
E_\mu (k)=
4t-2t\cos \left( \frac{\pi \mu}{N+1} \right) -2t\cos (k a)
\label{eq:cosBand}
\end{equation}
is identical to the Fermi energy $E_{\rm F}$.
The band edge of channel $\mu$ is defined by
$E_\mu (k=0)$. The band edge is located below
$E_{\rm F}$ for the conduction channel.
For $E_\mu (k=0)>\ef$, on the other hand, mode $\mu$ is
an evanescent mode. The wavefunction of the
evanescent mode is given by
\begin{equation}
\psi_\mu (i^\p , j^\p ) = \exp (-\kappa_{\mu}a j^\p)
u_{\mu} (i^\p),
\end{equation}
where $a j^\p$ is the distance from the junction along the lead and
$\kappa_{\mu}$ satisfies $E_\mu (i\kappa_{\mu})=\ef$.

\subsection{Calculation method}

For numerical studies, we introduce the Green's function.
The Green's function for the junction area is defined by
\begin{equation}
\hat{G}_{\sigma}(E) =
\left[ EI-\mathcal{H}_{\sigma}-\sum_p \Sigma^p \right]^{-1},
\label{eq:Green}
\end{equation}
where $\mathcal{H}_{\sigma}$ is the truncated matrix of the
Hamiltonian for the junction area ($N^2 \times N^2$) with
spin $\sigma$, and $\Sigma^p$ is the self-energy due to the
coupling to the lead $p$:
\begin{equation}
\Sigma^p =-t \ \tau_p^\da U \Lambda U^{-1} \tau_p^{\ }.
\end{equation}
Here $U$ is a unitary matrix,
$U=(\bm{u}_1, \bm{u}_2, \cdots, \bm{u}_N )$, where
$\bm{u}_{\mu}=\left[ u_{\mu}(1), u_{\mu}(2) , \cdots , u_{\mu}(N)
\right]^t$.
$\Lambda={\rm diag}(\lambda_1,\lambda_2,\cdots,\lambda_N)$,
where $\lambda_{\mu}=\exp (ik_{\mu} a)$ for conduction channels
and $\lambda_{\mu}=\exp (-\kappa_{\mu}a)$ for evanescent modes.
$\tau_p$ is a coupling matrix ($N \times N^2$) between
lead $p$ and the junction: $\tau_p (p_i,i)=1$ for $p_i$ being
an adjacent site in lead $p$ to site $i$ in the junction;
$\tau_p (p_i,i)=0$ otherwise.~\cite{Datta}

The spin-dependent conductance from reservoir $p$ to reservoir
$q$ is obtained from the Landauer-B\"{u}ttiker formula.
It is written as
\begin{equation}
G^{qp}_{\sigma}
=\frac{e^2}{h} \text{Tr} \left[ \Gamma^q \hat{G}_{\sigma}(E)
\Gamma^p \hat{G}_{\sigma}^\da(E) \right],
\label{eq:Conductance}
\end{equation}
where
\begin{equation}
\Gamma^p =i[\Sigma^p -{\Sigma^p}^\da].
\end{equation}
The total conductance is $G^{qp}=G^{qp}_++G^{qp}_-$, whereas
the spin polarization in the $z$ direction is defined by
\begin{equation}
P_z^{\ qp}= \frac{G^{qp}_+ - G^{qp}_-}{G^{qp}_+ + G^{qp}_-}
\label{eq:Polarization}
\end{equation}
for the current from reservoir $p$ to $q$.

To elucidate the virtual bound states in the potential well,
we calculate the DOS in the junction area. It is evaluated from
the Green's function (\ref{eq:Green}) as~\cite{com1}
\begin{equation}
D(E)=-\frac{1}{\pi} \sum_{\sigma} \text{Im}\left[\text{Tr}
\hat{G}_{\sigma}(E) \right].
\label{eq:DOS}
\end{equation}

%------------------------------
We assume that $\tilde{\lambda} =0.1$ for the strength of
SO interaction, which corresponds to the value for InAs,
$\lambda =117.1\, \mathrm{\AA^2}$,~\cite{Winkler}
with $a=W/30$ and width of the leads
$W \approx 50\,\mathrm{nm}$. The temperature $T=0$.
We focus on the transport from reservoir 1 to 2 and
omit the superscript 21 of $G_{\pm}^{21}$ and $P_z^{21}$,
otherwise stated. 
Note that the conductance from reservoir 1 to 3 is related to
$G_{\pm }^{31} =G_{\mp }^{21}$ from the symmetry of the system,
for both three- and four-terminal devices. The current from
reservoir 1 to 4 is not spin-polarized in the four-terminal device.
%------------------------------

%%%%%figure
\begin{figure}
\begin{center}
\includegraphics[width=7cm]{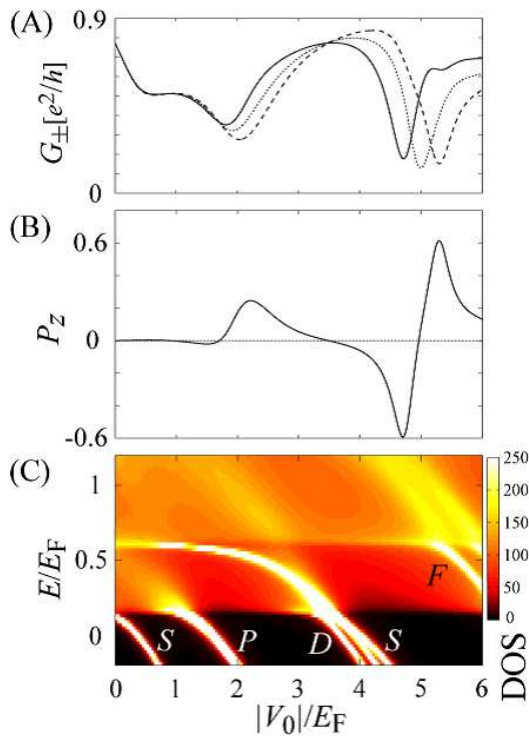}
\end{center}
\caption{(Color online)
Numerical results for the three-terminal device with
$\kf R_0 =2$, where $R_0$ is the radius of attractive
potential. (A) Conductance $G_{\pm}$ from reservoir 1
to 2 in Fig.\ \ref{fig:System}(A) for $s_z=\pm 1/2$ and
(B) spin polarization $P_z$ of the output current
in reservoir 2, as functions of the potential depth $|V_0|$.
In (A), solid and broken lines indicate $G_{+}$ and
$G_{-}$, respectively. A dotted line shows the
conductance per spin in the absence of the SO interaction.
(C) Grayscale plot of the density of states in the
junction area, $D(E)$, in the plane of $|V_0|$ and energy
$E$ of electron.}
\label{fig:T2}
\end{figure}
%%%%%%%%%%%%%%

\section{CALCULATED RESULTS}

We calculate the conductance $G_{\pm}$ for spin $s_z=\pm 1/2$
and spin polarization $P_z$ when the potential depth $|V_0|$
is tuned. We examine three cases of $\kf R_0 =1$, $2$, and
$3$, where the Fermi wavenumber $\kf$ is defined by the
Fermi energy $\ef$ as $\ef /t =(\kf a)^2$.\cite{com2}
In the three cases, the Fermi energy $\ef$ is different,
whereas the radius of the potential well $R_0$ is fixed.
The number of conduction channels in the leads is 1, 2, and 3,
respectively.

Here we discuss the cases of $\kf R_0 =2$ and $3$.
The numerical result with $\kf R_0 =1$ is given in Appendix B.
(Surprisingly, we find a perfect spin polarization $P_z=1$
in the case of $\kf R_0 =1$. However, the transport properties
seem quite specific. This is due to a strong interference effect
in the case of single conduction channel.)

\subsection{Case of $\kf R_0 =2$}

We begin with the three-terminal device in the
presence of two conduction channels in the leads
($\kf R_0 =2$). Figures \ref{fig:T2}(A) and \ref{fig:T2}(B) show the
conductance $G_{\pm}$ for $s_z=\pm 1/2$ and spin polarization
$P_z$, respectively,
when the potential depth $|V_0|$ is gradually changed.
As seen in Fig.\ \ref{fig:T2}(A),
the conductance $G_{\pm}$ shows three minima as a function of
$|V_0|$. At the first minimum at $|V_0|/\ef \approx 0.6$, the
difference in the conductance for $s_z=\pm 1/2$ is small.
At the second and third minima at
$|V_0|/\ef \approx 2$ and $5$, respectively,
the difference is remarkable, which results in a large spin
polarization in the $z$ direction [Fig.\ \ref{fig:T2}(B)].
$P_z$ is enhanced to 25\% around the second minimum and
61\% around the third minimum.

The behavior of $G_{\pm}$ should be ascribable to resonant
scattering at the potential well.
The resonant scattering takes place through a virtual bound state
in the potential well, which enhances the electron
scattering to the unitary limit (Appendix A).
This makes the minima of $G_{\pm}$ in our situation of both
three- and four-terminal devices.
(It is not trivial whether the resonant scattering makes minimum
or maximum of the conductance in multi-terminal devices.
See the discussion in Appendix A.) Around the minima of the
conductance, the difference between $G_{+}$ and $G_{-}$ is
magnified. Look at the third minimum of the conductance
around $|V_0|/\ef \approx 5$. The minimum of $G_{+}$ is
located at a smaller value of $|V_0|$ than that of $G_{-}$.
In consequence, $P_z$ shows a pair of negative dip
($G_{+}<G_{-}$) and a positive peak ($G_{+}>G_{-}$).
This dip-peak structure of $P_z$ can be understood in terms of
phase shifts when the resonance for $s_z=\pm 1/2$ is well
separated from each other (Appendix A).

%%%%%%%%%%%%%%
\begin{figure}
\begin{center}
\includegraphics[width=7cm]{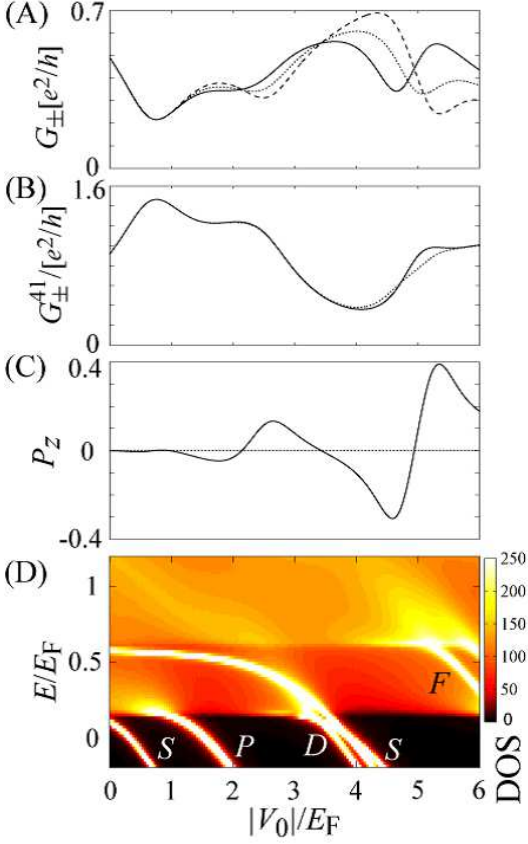}
\end{center}
\caption{(Color online)
Numerical results for the four-terminal device with
$\kf R_0 =2$, where $R_0$ is the radius of attractive
potential. (A) Conductance $G_{\pm}$ from reservoir 1
to 2 in Fig.\ \ref{fig:System}(B) for $s_z=\pm 1/2$,
(B) conductance $G_{\pm}^{41}$ from reservoir 1 to 4, and
(C) spin polarization $P_z$ of the output current
in reservoir 2, as functions of the potential depth $|V_0|$.
In (A) and (B), solid and broken lines indicate $G_{+}$ and
$G_{-}$, or $G_{+}^{41}$ and $G_{-}^{41}$, respectively.
A dotted line shows the
conductance per spin in the absence of the SO interaction.
In (B), solid and broken lines for $G_{\pm}^{41}$
are completely overlapped.
(D) Grayscale plot of the density of states in the
junction area, $D(E)$, in the plane of $|V_0|$ and energy
$E$ of electron.}
\label{fig:C2}
\end{figure}
%%%%%%%%%%%%%%

To confirm the above-mentioned scenario regarding the resonant
scattering, we calculate the DOS in the
junction area. Figure \ref{fig:T2}(C) shows a grayscale
plot of the DOS $D(E)$ in the plane of $|V_0|$ and energy $E$ of
electron. The band edge for the lowest and second conduction
channels in the leads [$E_1 (k=0)$ and $E_2 (k=0)$ in
Eq.\ (\ref{eq:cosBand})] are located at $E/\ef =0.154$ and $0.615$,
respectively. The sharp peaks of $D(E)$ below
the lowest band edge correspond to the bound states
inside the junction area. With an increase in the potential
depth $|V_0|$, several bound states appear one after another.
The first one is an $S$-like bound state ($l_z=0$)
although the angular momentum $l_z$ is not a good quantum
number in our device because of the lack of rotational
symmetry. The bound state
exists even without the potential well ($|V_0|=0$) in the
junction area,~\cite{Kiselev} and changes to the
$S$-like bound state in the potential well with increasing
$|V_0|$. The state is doubly degenerate due to the Kramers
degeneracy. The next are $P$-like bound states ($l_z=\pm 1$).
They are a pair of Kramers degenerate states.
Then $D$-like bound states ($l_z=\pm 2$) appear, which
are clearly split into two by the SO interaction.
Another $S$-like state is located at approximately the
same energy. Finally $F$-like bound states ($l_z=\pm 3$)
appear in Fig.\ \ref{fig:T2}(C).
The pair of Kramers degenerate states is largely
separated for the $F$-like states.

The peaks of the bound states in $D(E)$ are broadened above the
band edge of the lowest conduction channel in the leads,
which significantly influence
the electron scattering at the Fermi level as virtual bound states.
The second and third minima of the conductance $G_{\pm}$
are located at the position of $D$ and $F$-like virtual bound
states at $\ef$, respectively. This is a clear evidence of the resonant
scattering through virtual bound states. (At the first minimum of
$G_{\pm}$ around $|V_0|/\ef = 0.6$, we cannot find
any virtual bound state at the Fermi level. The minimum of
$G_{\pm}$ may not be due to the resonant scattering, but due to
some interference effect around the junction.)

We present the calculated results for the four-terminal device
with $\kf R_0 =2$ in Fig.\ \ref{fig:C2}:
(A) conductance $G_{\pm}$ for $s_z=\pm 1/2$ from
reservoir 1 to 2 in Fig.\ \ref{fig:System}(B),
(B) conductance $G_{\pm}^{41}$ from reservoir 1 to 4,
and (C) spin polarization $P_z$ of the output current
in reservoir 2, as functions of the potential depth $|V_0|$.
As seen in Fig.\ \ref{fig:C2}(B), $G_{+}^{41}=G_{-}^{41}$
because the SHE does not make a spin polarization in the
output current in reservoir 4.
The characters of conductance $G_{\pm}$
for $s_z=\pm 1/2$ and spin polarization
$P_z$ are almost the same as those in Fig.\ \ref{fig:T2}
for three-terminal device.
The conductance shows three minima. The second and third
minima are clearly due to resonant
scattering via $D$- or $F$-like virtual bound states,
as seen in the DOS in Fig.\ \ref{fig:C2}(D).
Around the minima, the conductance for $s_z=\pm 1/2$
is largely split by the SO interaction, which results
in a large spin polarization $P_z$.

%%%%figure
\begin{figure}
\begin{center}
\includegraphics[width=7cm]{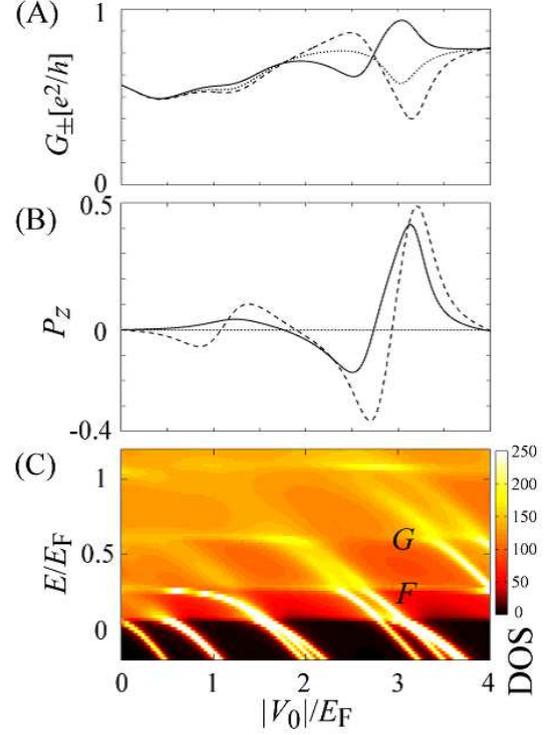}
\end{center}
\caption{(Color online)
Numerical results for the three-terminal device with
$\kf R_0 =3$, where $R_0$ is the radius of attractive
potential. (A) Conductance $G_{\pm}$ from reservoir 1
to 2 in Fig.\ \ref{fig:System}(A) for $s_z=\pm 1/2$ and
(B) spin polarization $P_z$ of the output current
in reservoir 2, as functions of the potential depth $|V_0|$.
In (A), solid and broken lines indicate $G_{+}$ and
$G_{-}$, respectively. A dotted line shows the
conductance per spin in the absence of the SO interaction.
(C) Grayscale plot of the density of states in the
junction area, $D(E)$, in the plane of $|V_0|$ and energy
$E$ of electron.
Regarding the result for four-terminal device with
$\kf R_0 =3$, a broken line in (B) indicates the
spin polarization $P_z$ of the output current in reservoir 2
in Fig.\ \ref{fig:System}(B).}
\label{fig:TC3}
\end{figure}
%%%%%%%%%%%%%%

\subsection{Case of $\kf R_0 =3$}

Figure \ref{fig:TC3} shows the calculated results for
the three-terminal device in the presence of
three conduction channels in the leads ($\kf R_0 =3$):
(A) conductance $G_{\pm}$ for $s_z=\pm 1/2$ and
(B) spin polarization $P_z$ in the $z$ direction,
as functions of the potential depth $|V_0|$.
Figure \ref{fig:TC3}(C) shows
a grayscale plot of the density of states $D(E)$ in the
plane of $|V_0|$ and energy $E$ of electron.
In Fig.\ \ref{fig:TC3}(B), a broken line indicates the
spin polarization $P_z$ in the four-terminal device.

The conductance $G_{\pm}$ shows several minima as a function of
potential depth $|V_0|$. The spin polarization $P_z$ is enhanced
around the minima of $G_{\pm}$. These properties can be
understood in the same way as in the preceding subsection.
The polarization $P_z$ is enhanced to $41$\% at $|V_0|/\ef = 3.1$
in the three-terminal device, and it is enhanced to $49$\%
at $|V_0|/\ef = 3.2$ in the four-terminal device.
This is due to resonant scattering via $G$-like virtual
bound states ($l_z=\pm 4$).

Compared with the case of two conduction channels
in Figs.\ \ref{fig:T2} and \ref{fig:C2}, the values of the conductance $G_{\pm}$
are larger in the case of three conduction channels
($\kf R_0 =3$), whereas
the maximum value of spin polarization is almost the same.
This implies a more efficient spin filter in the case of
three conduction channels than in the case of two conduction
channels.

\subsection{Channel analysis for spin filtering}

%%%%figure
\begin{figure}
\begin{center}
\includegraphics[width=6cm]{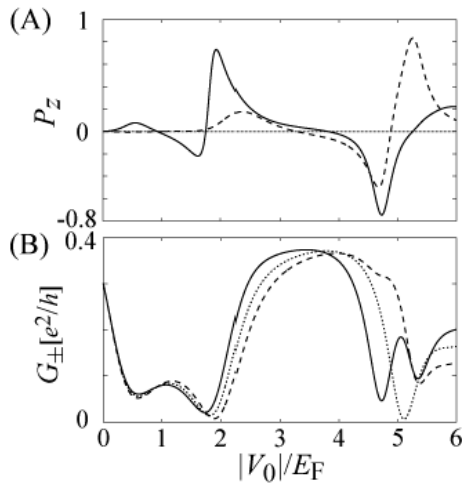}
\end{center}
\caption{
A channel analysis for incident waves from reservoir 1
in the three-terminal device with $\kf R_0 =2$.
(A) Spin polarization $P_z$ of the output current
in reservoir 2 in Fig.\ \ref{fig:System}(A),
as a function of the potential depth $|V_0|$,
for the incident electrons in the lowest channel (solid
line) and second channel (broken line).
(B) Conductance $G_{\pm}$ from reservoir 1 to 2
for the incident electrons in the lowest channel
with $s_z=1/2$ (solid line) and $-1/2$ (broken line).
A dotted line indicates the conductance per spin in the
absence of the SO interaction.}
\label{fig:ch2}
\end{figure}
%%%%%%%%%%%%%%

In cases of $\kf R_0 =2$ and $3$,
there are two and three conduction channels in the leads,
respectively. To examine the resonant scattering in detail,
we perform a channel analysis of incident waves from reservoir 1.
The results are given only for the three-terminal device
in this subsection.

In the case of $\kf R_0 =2$, we plot the spin polarization
$P_z$ for the incident electrons in the lowest and second
channels in Fig.\ \ref{fig:ch2}(A). At $|V_0|/\ef \sim 2$
(resonance by $D$-like virtual bound state), $P_z$
is enhanced to 73\% for the lowest channel while 
it is to 18\% for the second channel. Hence the former
plays a main role in the spin polarization. At
$|V_0|/\ef \sim 5$ (resonance by $F$-like virtual bound state),
on the other hand, $|P_z|$ becomes 75\% for the lowest channel
while it becomes 83\% for the second channel. Then both
channels are important for the spin-dependent scattering.

We could selectively inject the lowest channel to the junction,
e.g., using a quantum point contact fabricated on the lead
connected to reservoir 1. Then we could realize a spin filter
with an efficiency of about 75\%. In Fig.\ \ref{fig:ch2}(B),
we plot the conductance $G_{\pm}$ when only the lowest channel
is injected from reservoir 1.
At $|V_0|/\ef \sim 2$, the conductance almost vanishes
although $P_z$ is enhanced to 73\%. This is due to an interference
effect at the junction as in the case of
single conduction channel with $\kf R_0 =1$ (Appendix B).
At $|V_0|/\ef \sim 5$, on the other hand, the total conductance is
$G_++G_-=0.4 (e^2/h)$ and $P_z=-75$\%.
The latter situation is favorable to application to a
spin filter.

\begin{figure}
\begin{center}
\includegraphics[width=6cm]{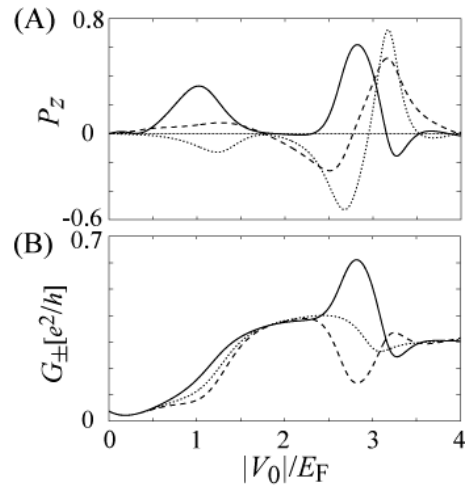}
\end{center}
\caption{A channel analysis for incident waves from reservoir 1
in the three-terminal device with $\kf R_0 =3$.
(A) Spin polarization $P_z$ of the output current
in reservoir 2 in Fig.\ \ref{fig:System}(A),
as a function of the potential depth $|V_0|$,
for the incident electrons in the lowest channel (solid
line), second channel (broken line), and third
channel (dotted line).
(B) Conductance $G_{\pm}$ from reservoir 1 to 2
for the incident electrons in the lowest channel
with $s_z=1/2$ (solid line) and $-1/2$ (broken line).
A dotted line indicates the conductance per spin in the
absence of the SO interaction.}
\label{fig:ch3}
\end{figure}

A similar channel analysis is given for the case of $\kf R_0 =3$
in Fig.\ \ref{fig:ch3}. There are three incident channels
in this case.
It is notable that, at $|V_0|/\ef \sim 2.8$,
a spin polarization of $P_z=62$\% is realized for the incident
electrons in the lowest channel while the total
conductance is as large as $G_++G_-=0.8 (e^2/h)$.

%%%%figure
%%%%%%%%%%%%%%

\subsection{Repulsive potential}

\begin{figure}
\begin{center}
\includegraphics[width=6cm]{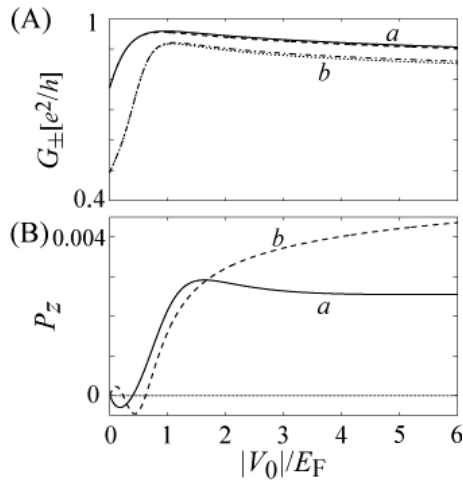}
\end{center}
\caption{Numerical results for the three-terminal device
(curves labeled by $a$) and four-terminal device
%%%%%%%
(curves labeled by $b$) with repulsive potential [$V_0>0$ in
Eq.\ (\ref{eq:potential})]. $\kf R_0 =2$, where $R_0$ is
the radius of the potential.
(A) Conductance $G_{\pm}$ from reservoir 1
to 2 in Fig.\ \ref{fig:System}, as a function of potential
height $V_0$, for $s_z=1/2$ (solid lines) and $-1/2$ (broken lines).
(B) Spin polarization $P_z$ of the output current
in reservoir 2.
}
\label{fig:R}
\end{figure}

We investigate the SHE caused by the scattering by
a repulsive potential, $V_0>0$ in Eq.\ (\ref{eq:potential}).
Figure \ref{fig:R} shows (A) conductance $G_{\pm}$ for
$s_z=\pm 1/2$ and (B) spin polarization $P_z$ in the $z$ direction,
when the potential height $V_0$ is gradually increased.
The extrinsic SHE is expected even with a repulsive
potential.~\cite{Yamamoto} However, the spin-filtering effect
is much weaker than the case with an
attractive potential. In Fig.\ \ref{fig:R}(B), the
spin polarization is at most $P_z \approx 0.3\%$ in the three-terminal
device and $P_z \approx 0.45\%$ in the four-terminal device. In this
case, the resonant scattering does not take place since
virtual bound states are hardly formed in the potential
barrier. This indicates an important role of resonant
scattering in the enhancement of the SHE discussed in the
preceding subsections.

%%%%figure
%%%%%%%%%%%%%%

\section{CONCLUSIONS AND DISCUSSION}

We have numerically studied the extrinsic SHE in
multi-terminal devices including an antidot, fabricated
on semiconductor heterostructures with strong SO interaction.
The antidot creates a tunable potential on two-dimensional
electron gas in the heterostructures, which may be attractive
as well as repulsive. When an attractive potential
is tuned properly, the resonant scattering via a virtual
bound state takes place, which makes minima of the
conductance from reservoir 1 to 2 in Fig.\ \ref{fig:System}.
Then the difference between the conductances for
$s_z=\pm 1/2$ is enlarged and, as a result, the spin
polarization is significantly enhanced in the direction
perpendicular to the two-dimensional plane.
The spin polarization can be more than 50\%
in our three- and four-terminal devices.

The enhancement of the extrinsic SHE by resonant scattering
has been studied in different systems.
Kiselev and Kim
have proposed a three-terminal spin filter without antidot
in the presence of Rashba SO interaction
[Eq.\ (\ref{eq:Rashba})].~\cite{Kiselev,Kiselev2}
They have pointed out that the spin-filtering effect
is enhanced by resonant scattering at the junction area
when the Fermi energy of 2DEG is tuned.
In their device, the direction of spin polarization is tilted
from the $z$ direction perpendicular to the plane.
In our device, the spin is polarized in the $z$ direction,
which is easier to detect by an optical experiment on the
Kerr rotation~\cite{Kato} and, above all, more suitable to
spintronic devices.

The extrinsic SHE enhanced by (many-body) resonant scattering
has been examined for metallic systems with magnetic
impurities.~\cite{Fert,Fert2,Guo} In the case of
semiconductor heterostructures, however, we have a great
advantage in the tunability of scattering potential.
The enhanced SHE caused by resonant scattering at a
single potential can be investigated in details.

We make some comments regarding our calculations.
(i) The electron-electron interaction has been neglected.
Below the band edge of the lowest conduction channel in
the leads,
we have observed bound states in the density of states
in the potential well.
Since the bound states are occupied by electrons, we
have to consider the electron-electron interaction
between the electrons and conduction electrons at the Fermi
level. The Hartree potential from the trapped electrons
should be taken into account
although the Coulomb blockade is irrelevant to the case of antidot
potential without tunnel barriers, in contrast to
the case of conventional quantum dots.~\cite{Kouwenhoven}
In our calculated results, therefore, the
values of $|V_0|$ at the resonance are underestimated.

(ii) It is necessary to create such a deep potential as
$|V_0| \sim E_{\rm F}$ in designing the devices.
This might be difficult with a usual antidot structure
fabricated on semiconductor heterostructures.
Alternatively, we could make such a deep potential using
a STM tip, a charged impurity under the antidot, etc.

(iii) We have assumed that the antidot potential $V(\bm{r})$
is independent of $z$. Otherwise, the Rashba-type SO
interaction, Eq.\ (\ref{eq:Rashba}) with
$\alpha=\lambda(\partial V/\partial z)$,
must be added to Eq.\ (\ref{eq:SO2D}).
This would create an effective magnetic field in the $xy$ plane
and thus decrease the spin polarization in the $z$ direction.
The Dresselhaus SO interaction has also been disregarded.
The SO interaction is induced by the inversion asymmetry of
the crystal~\cite{Dresselhaus} and expressed as
\begin{equation}
H_{\rm SO}=\frac{\beta}{\hbar}(-p_x\sigma_x+p_y\sigma_y).
\end{equation}
This would also result in an effective magnetic field in the
$xy$ plane and lessen the spin polarization in the
$z$ direction.

\section*{ACKNOWLEDGMENTS}

This work was partly supported by the Strategic Information
and Communications R\&D Promotion Program (SCOPE) from the
Ministry of Internal Affairs and Communications of Japan, and
by a Grant-in-Aid for Scientific Research from
the Japan Society for the Promotion of Science.

%%%%%%%%%%%%%%%%
\appendix
\section{Formulation of Spin Hall Effect in 2DEG}

Here, we summarize our previous study in Ref.\ \onlinecite{Eto}.

First, we give a quantum mechanical formulation of the
extrinsic SHE for 2DEG.
For the scattering problem with Eq.\ (\ref{eq:Schroedinger}),
we adopt the partial wave expansion with
$l_z=m=0$, $\pm 1$, $\pm 2$, $\cdots$.~\cite{Aharonov-Bohm}
As an incident wave, we consider a plane wave propagating in
the $x$ direction, $e^{ikx}$, with spin $s_z=1/2$ or $-1/2$.
$E=\hbar^2 k^2/(2m^*)$. The plane wave is expanded as
\begin{equation}
e^{ikx}=e^{ikr \cos \theta}=\sum_{m=-\infty}^{\infty}
i^m J_{m}(kr) e^{im\theta},
\label{eq:incident}
\end{equation}
where $\theta$ is the angle from the $x$ direction and
$J_{m}$ is the $m$th Bessel function. Its asymptotic form at
$r \rightarrow \infty$ is given by
$J_{m}(kr)\sim \sqrt{2/(\pi kr)} \cos(kr-m\pi/2-\pi/4)$.
In the solution of Eq.\ (\ref{eq:Schroedinger}),
$J_{m}(kr)$ in
Eq.\ (\ref{eq:incident}) is replaced by $R_m^{\pm}(r)$ for
$s_z=\pm 1/2$,
\begin{equation}
\psi^\pm({\bm r}) = \sum_{m=-\infty}^{\infty} i^m R_m^{\pm}(r) e^{im\theta}
\label{eq:solution1}
\end{equation}
for spin $s_z=\pm 1/2$, where $R_m^{\pm}(r)$ satisfies
\[
\left[
-\frac{\hbar^2}{2m^*} \left(
\frac{d^2}{dr^2}+\frac{1}{r}\frac{d}{dr}-\frac{m^2}{r^2}
\right)+V(r) \pm \frac{m}{2} V_1(r)
\right] R_m^{\pm}(r) 
\]
\begin{equation}
= E R_m^{\pm}(r).
\label{eq:Req}
\end{equation}
Its asymptotic form determines the phase shift $\delta_m^{\pm}$:
\begin{equation}
R_m^{\pm}(r) \sim \sqrt{\frac{2}{\pi kr}} e^{i \delta_m^{\pm}}
\cos \left( kr-\frac{m\pi}{2}-\frac{\pi}{4}+\delta_m^{\pm}
\right).
\label{eq:Req2}
\end{equation}
From Eqs.\ (\ref{eq:Req}) and (\ref{eq:Req2}),
we immediately obtain the relation of
$\delta_{-m}^{\pm}=\delta_m^{\mp}$, indicating the time reversal
symmetry. The SO interaction is not relevant to the $S$ wave
($m=0$): $\delta_0^{+}=\delta_0^{-} \equiv \delta_0$.

The scattering amplitude $f^{\pm}(\theta)$
for spin $s_z=\pm 1/2$ is expressed in terms of phase shifts.
The asymptotic form of the
wavefunction in Eq.\ (\ref{eq:solution1}) is given by
\begin{equation}
\psi^\pm \sim
e^{ikx} + f^{\pm}(\theta) \frac{e^{i(kr+\pi/4)}}{\sqrt{r}},
\label{eq:asymptotic}
\end{equation}
where $f^{\pm}(\theta)$ is related to the $S$ matrix by
\begin{eqnarray}
f^{\pm}(\theta)
& = &
\sum_m f^{\pm}_m e^{im\theta},
\label{eq:scatt-ampl} \\
S_m^{\pm}
& = &
1+i\sqrt{2\pi k}f_m^{\pm}=e^{2i\delta_m^{\pm}}.
\label{eq:smatrix}
\end{eqnarray}
% From Eqs.\ (\ref{eq:scatt-ampl}) and (\ref{eq:smatrix}), 
From these equations, we obtain
\begin{equation}
f^{\pm}(\theta) = f_1(\theta) \pm f_2(\theta),
\label{eq:f0}
\end{equation}
where
\begin{eqnarray}
f_1(\theta) &=&
\frac{1}{i \sqrt{2\pi k}} \Biggl[
e^{2i\delta_0}-1
\nonumber \\
& & +
\sum_{m=1}^{\infty}
\left(e^{2i\delta_m^+} + e^{2i\delta_m^-}-2 \right)
\cos m\theta \Biggr],
\label{eq:f1}
\\
f_2(\theta) &=&
\frac{1}{\sqrt{2\pi k}} \sum_{m=1}^{\infty}
(e^{2i\delta_m^+}-e^{2i\delta_m^-})
\sin m\theta.
\label{eq:f2}
\end{eqnarray}

The scattering cross section is given by
$\sigma^{\pm}(\theta)=|f^{\pm}(\theta)|^2$.
Hence the spin polarization of the scattered wave in
the $\theta$ direction is expressed as
\begin{equation}
P_z=\frac{|f^+|^2-|f^-|^2}{|f^+|^2+|f^-|^2}
=\frac{2{\rm Re}(f_1 {f_2}^*)}{|f_1|^2+|f_2|^2},
\label{eq:eSHEpol}
\end{equation}
when the incident electron is unpolarized.
This formula is analogous to that of skew scattering in
three dimensions.~\cite{Mott,Landau}
The spin is polarized in the $z$ direction
and $P_z(-\theta)=-P_z(\theta)$.

A remark is made on Eq.\ (\ref{eq:asymptotic}).
We put a phase of $\pi/4$ on the exponent of the second term on
the right side. Although the phase has no physical meaning, it
ensures the ``optical theorem'' that the total cross section
is determined only by the amplitude of forward scattering;
\begin{equation}
\sigma^{\pm}_{\rm total} \equiv \int_0^{2\pi}
\sigma^{\pm}(\theta) d\theta=\sqrt{\frac{8\pi}{k}}
{\rm Im} f^{\pm}(0).
\end{equation}

%%%%figure
\begin{figure}
\begin{center}
\includegraphics[width=7cm]{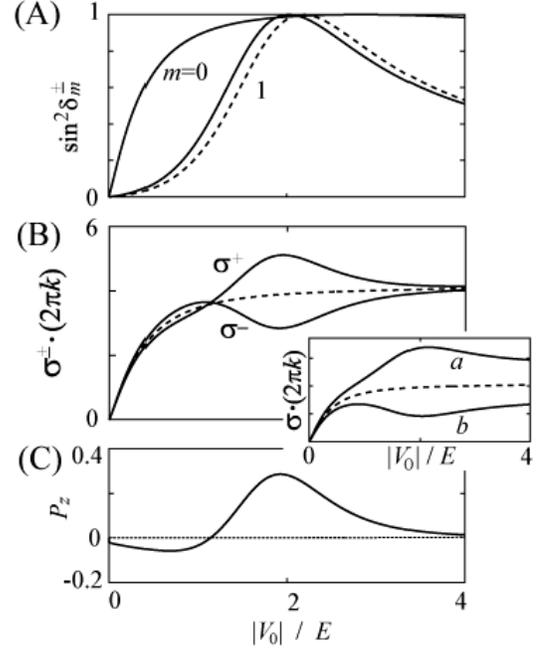}
\end{center}
\caption{Partial wave expansion for the
extrinsic SHE due to the scattering by
a potential well, Eq.\ (\ref{eq:pot-well}),
in 2DEG.\ $kR_0=1$.
The strength of the SO interaction is $\lambda k^2=0.01$.
(A) Scattering probability of each partial wave,
$\sin^2 \delta_m^{\pm}$,
(B) scattering cross section $\sigma^{\pm}(\theta=-\pi/2)$ for
$s_z=\pm 1/2$, and
(C) spin polarization $P_z$ at $\theta =-\pi/2$,
as functions of the potential depth $|V_0|$
[normalized by electron energy $E=\hbar^2 k^2/(2m^*)$].
In (A), solid and broken lines indicate the cases of
$s_z=1/2$ and $-1/2$, respectively, for $m=1$
($\delta_{-1}^{\pm}=\delta_1^{\mp}$). The scattering
probability for $|m| \ge 2$ is negligible.
In (B), a broken line indicates the cross section at
$\theta=-\pi/2$ in the absence of SO interaction.
Inset: scattering cross section in the absence of SO interaction
at $\theta=-0.45\pi$ ($a$), $-\pi/2$ (broken line),
and $\theta=-0.55\pi$ ($b$).}
\label{fig:appendA1}
\end{figure}
%%%%%%%%%%%%%%

Now we apply our formula of the SHE to the case of a potential well,
Eq.\ (\ref{eq:potential}) with $\Delta R_0 \rightarrow 0$, that is,
\begin{equation}
V(r)=V_0\theta(R_0-r)
\label{eq:pot-well}
\end{equation}
($V_0<0$), where $\theta(t)$ is a step function [$\theta(t)=1$
for $t>0$, $0$ for $t<0$].
Then $V_1=-(2 \lambda/r)V'(r)=(2\lambda/R_0) V_0 \delta(r-R_0)$
with the $\delta$ function $\delta(t)$.

The phase shifts $\delta_m^{\pm}$ are calculated by solving
Eq.\ (\ref{eq:Req}). The calculation is elementary.
We have only to consider the case with $m \ge 0$ because of
the relation of $\delta_{-m}^{\pm}=\delta_{m}^{\mp}$. For
$r>R_0$, $V=0$ and thus
\begin{eqnarray}
R_m^{\pm}(r) & = &
C_1 J_m(kr) + C_2 Y_m(kr)
\label{eq:appendB2}
\\
& \sim &
\sqrt{\frac{2}{\pi kr}} [
C_1 \cos(kr-m\pi/2-\pi/4) 
\nonumber \\
& &
+ C_2 \sin(kr-m\pi/2-\pi/4)],
\label{eq:appendB3}
\end{eqnarray}
where $Y_m$ is the $m$th Neumann function. From Eqs.\
(\ref{eq:appendB3}) and (\ref{eq:Req2}), we find
\begin{equation}
\tan \delta_m^{\pm} = -C_2/C_1.
\label{eq:appendB4}
\end{equation}
For $r<R_0$, $V=V_0$. Then
\begin{eqnarray}
R_m^{\pm}(r) & = & C_3 J_m(k'r),
\label{eq:appendB5}
\\
\hbar^2 k'^2/(2m^*) & = & E-V_0.
\end{eqnarray}
The connection of Eqs.\ (\ref{eq:appendB2}) and (\ref{eq:appendB5})
at $r=R_0$ yields
\begin{equation}
\tan \delta_m^{\pm} =
\frac{[J_{m-1}(kR_0)-J_{m+1}(kR_0)]J_{m}(k'R_0)
-\alpha_m^{\pm} J_{m}(kR_0)}
{[Y_{m-1}(kR_0)-Y_{m+1}(kR_0)]J_{m}(k'R_0)-
\alpha_m^{\pm} Y_{m}(kR_0)}
\end{equation}
with
\begin{eqnarray}
\alpha_m^{\pm} =
(k'/k)[J_{m-1}(k'R_0)-J_{m+1}(k'R_0)] 
\nonumber \\
\mp
2m [1+(k'/k)^2](k\lambda/R_0) J_{m}(k'R_0).
\end{eqnarray}
Here, we have used the relation of
$dJ_m(x)/dx=[J_{m-1}(x)-J_{m+1}(x)]/2$
and $dY_m(x)/dx=[Y_{m-1}(x)-Y_{m+1}(x)]/2$.

%%%%figure
\begin{figure}
\begin{center}
\includegraphics[width=6cm]{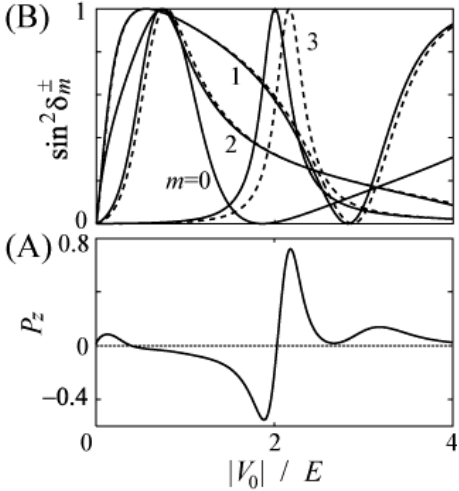}
\end{center}
\caption{
Partial wave expansion for the
extrinsic SHE due to the scattering by
a potential well, Eq.\ (\ref{eq:pot-well}),
in 2DEG.\ $kR_0=2$.
The strength of the SO interaction is $\lambda k^2=0.01$.
(A) sScattering probability of each partial wave,
$\sin^2 \delta_m^{\pm}$, and
(B) spin polarization $P_z$ at $\theta =-\pi/2$,
as functions of the potential depth $|V_0|$
[normalized by electron energy $E=\hbar^2 k^2/(2m^*)$].
In (A), solid and broken lines indicate the cases of
$s_z=1/2$ and $-1/2$, respectively, for $m>0$
($\delta_{-m}^{\pm}=\delta_{m}^{\mp}$).
}
\label{fig:appendA2}
\end{figure}
%%%%%%%%%%%%%%

The calculated results are shown in Fig.\ \ref{fig:appendA1}:
(A) scattering probability of each partial
wave, $\sin^2 \delta_m^{\pm}$, (B) scattering cross section
$\sigma^{\pm}(\theta=-\pi/2)$ for $s_z=\pm 1/2$, and
(C) spin polarization $P_z$ at $\theta=-\pi/2$, when
the potential depth $|V_0|$ is gradually changed.
The strength of the SO interaction is set to be
$\lambda k^2=0.01$, which corresponds to the
value for InAs, $\lambda=117.1$
\AA$^2$,~\cite{Winkler} with electron wavelength
$2\pi/k=70$ nm. The radius of the potential well is $R_0=1/k$.
As seen in Fig.\ \ref{fig:appendA1}(A),
with an increase in $|V_0|$, the scattering probability
increases and becomes unity at some
values of $|V_0|$ (unitary limit with $\delta_m^{\pm}=\pi/2$) for
$m=0$ ($S$ wave) and $m=\pm 1$ ($P$ wave). This is due to
resonant scattering through virtual bound states in the potential
well. The resonant width is narrower for larger $|m|$ because
of the centrifugal potential $\propto m^2/r^2$ separating the
bound states from the outer region.

As discussed in the text with Eq.\ (\ref{eq:Schroedinger}),
the scattering of a partial wave with positive $m$ is
enhanced (suppressed) for spin $s_z=1/2$ ($s_z=-1/2$).
Around the resonance of the $P$ waves,
the scattering of $(m,s_z)=(1,1/2)$
goes to the unitary limit at a smaller
value of $|V_0|$ than the scattering of $(m,s_z)=(1,-1/2)$.
Thus we observe a difference between
$\delta_1^{+}$ and $\delta_1^{-}$ in 
Fig.\ \ref{fig:appendA1}(A).
This leads to the difference of the scattering cross section
at $\theta=-\pi/2$ for $s_z=\pm 1/2$
[Fig.\ \ref{fig:appendA1}(B)] and
a spin polarization [$P_z \approx 30\%$ in
Fig.\ \ref{fig:appendA1}(C)].
When $\delta_m^{\pm}$ ($|m| \ge 2$) is negligible,
Eqs.\ (\ref{eq:f0})-(\ref{eq:f2}) yield
\begin{eqnarray}
\sigma^{\pm}(\theta=-\pi/2)
=\frac{2}{\pi k} [\sin^2 \delta_0+\sin^2 \Delta \delta_1
\nonumber \\
\pm 2 \sin \delta_0 \sin(2\bar{\delta}_1-\delta_0)
\sin \Delta \delta_1],
\end{eqnarray}
where $\Delta \delta_1=\delta_1^{+}-\delta_1^{-}$ and
$\bar{\delta}_1=(\delta_1^{+}+\delta_1^{-})/2$.
Around the resonance, $\Delta \delta_1$ is enlarged,
which results in the enhancement of the SHE.
For $\bar{\delta}_1\approx \pi/2$, the spin polarization
is given by
\begin{equation}
P_z(\theta=-\pi/2) \approx
\frac{2 \sin^2 \delta_0 \sin \Delta \delta_1}
{\sin^2 \delta_0 + \sin^2 \Delta \delta_1}
\end{equation}
from Eq.\ (\ref{eq:eSHEpol}).

%%%%figure
\begin{figure}
\begin{center}
\includegraphics[width=7cm]{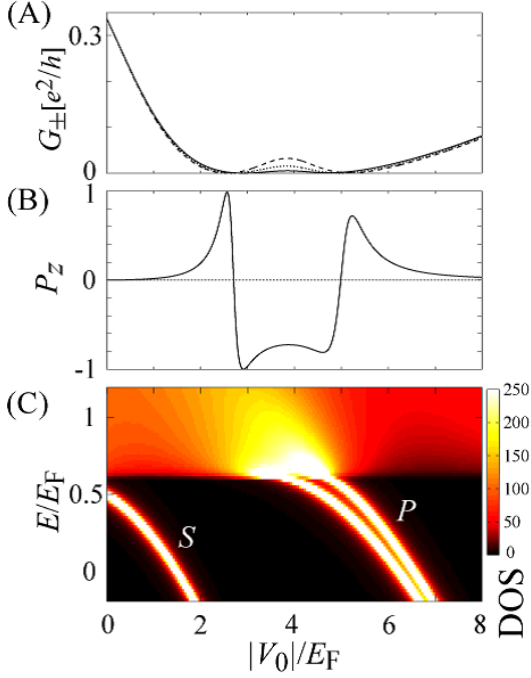}
\end{center}
\caption{(Color online)
Numerical results for the three-terminal device with
$\kf R_0 =1$, where $R_0$ is the radius of attractive
potential. (A) Conductance $G_{\pm}$ from reservoir 1
to 2 in Fig.\ \ref{fig:System}(A) for $s_z=\pm 1/2$ and
(B) spin polarization $P_z$ of the output current
in reservoir 2, as functions of the potential depth $|V_0|$.
In (A), solid and broken lines indicate $G_{+}$ and
$G_{-}$, respectively. A dotted line shows the
conductance per spin in the absence of the SO interaction.
(C) Grayscale plot of the density of states in the
junction area, $D(E)$, in the plane of $|V_0|$ and energy
$E$ of electron.
}
\label{fig:T1}
\end{figure}
%%%%%%%%%%%%%%

In our three- and four-terminal devices, the resonant scattering
makes minima of the conductance. It should be hard to say,
however, whether the resonant scattering enhances or suppresses
the conductance in general.
For simplicity, let us ignore the SO interaction.
We plot the scattering cross section in the inset in Fig.\
\ref{fig:appendA1}(B), in the direction of $\theta=-0.45\pi$,
$-\pi/2$, and $\theta=-0.55\pi$. Using $\delta_0$ and
$\delta_1^{+}=\delta_1^{-} \equiv \delta_1$, the cross
section is written as
\begin{eqnarray}
\sigma(\theta)=\frac{2}{\pi k}
[\sin^2 \delta_0+4\sin^2 \delta_1 \cos^2 \theta
\nonumber \\
+4\sin \delta_0 \sin \delta_1 \cos (\delta_0-\delta_1) \cos \theta].
\end{eqnarray}
The resonance of the $P$ wave ($\delta_1=\pi/2$) has no
effect on the scattering at $\theta=-\pi/2$. Its
effect on the scattering at $\theta \ne -\pi/2$
depends on the value of $\delta_0$.
In the inset in Fig.\ \ref{fig:appendA1}(B), the resonance of
the $P$ wave makes a peak of $\sigma(\theta=-0.45\pi)$
and a dip of $\sigma(\theta=-0.55\pi)$.
In the situation of our devices, the conductance should reflect
an integrated value of $\sigma(\theta)$ around $\theta=-\pi/2$,
and also a complicated interference effect.

Figure \ref{fig:appendA2} shows the calculated results in the
case of $kR_0=2$. The resonant scattering is observed for
$0 \le |m| \le 3$. Around the resonance of $F$ waves ($|m|=3$), 
$P_z$ is enhanced to 72\%. In general, a sharper resonance
enlarges $\delta_{m}^{+}-\delta_{m}^{-}$ for larger $|m|$,
which results in a larger polarization.

In Fig.\ \ref{fig:appendA2}(B), $P_z$ shows a dip and peak
structure at the $F$-wave resonance.
This structure is observed when the resonance of 
$(m,s_z)=(3,1/2)$ is sufficiently separated from that of
$(m,s_z)=(3,-1/2)$.
Around the $F$-wave resonance, 
$\delta_1^+ \approx \delta_1^-$ and
$\delta_2^+ \approx \delta_2^-$ ($\equiv \delta_2$), as seen
in Fig.\ \ref{fig:appendA2}(A). Neglecting $\delta_0$,
Eq.\ (\ref{eq:eSHEpol}) yields
\begin{equation}
P_z(\theta=-\pi/2) \approx
\frac{4 \sin \delta_2 \sin(2\bar{\delta}_3-\delta_2)
\sin \Delta \delta_3}
{4\sin^2 \delta_2 + \sin^2 \Delta \delta_3},
\end{equation}
where $\Delta \delta_3=\delta_3^{+}-\delta_3^{-}$ and
$\bar{\delta}_3=(\delta_3^{+}+\delta_3^{-})/2$.
At the resonance of $(m,s_z)=(3,1/2)$, $\delta_3^{+}=\pi/2$
and $\delta_3^{-} \approx 0$. Then $P_z(\theta=-\pi/2)$
shows a dip for $\pi/2 < \delta_2 < \pi$.
At the resonance of $(m,s_z)=(3,-1/2)$, $\delta_3^{+}
\approx \pi$ and $\delta_3^{-} =\pi/2$. Then
$P_z(\theta=-\pi/2)$ shows a peak.
A similar dip-peak structure of $P_z$ is observed
around $F$-wave resonance in Figs.\ \ref{fig:T2} and
\ref{fig:C2} for our devices.

\section{Case of $\kf R_0 =1$}

We present calculated results in the case of
attractive potential and $\kf R_0 =1$, where the number of
conduction channels is unity in the leads. In this case, an
interference effect around the junction strongly
influences the conductance and spin polarization.
We discuss the results only for the three-terminal device.

Figures \ref{fig:T1}(A) and \ref{fig:T1}(B) show the conductance $G_{\pm}$
for $s_z=\pm 1/2$ and spin polarization $P_z$, respectively,
when the potential depth $|V_0|$ is gradually changed.
As seen in Fig.\ \ref{fig:T1}(A), the
conductance $G_{\pm}$ vanishes two times at
$|V_0|/\ef \approx 2.7$ and $5.0$.
The reason why the conductance completely disappears
cannot be explained by resonant scattering only.
This is due to an interference effect around the
junction area. The value of $|V_0|$ for $G_{+}=0$ and
that for $G_{-}=0$ are different from each other, in the
presence of the SO interaction. As a result,
$P_z=1$ at $G_{-}=0$ and
$P_z=-1$ at $G_{+}=0$, as seen in Fig.\ \ref{fig:T1}(B).

Figure \ref{fig:T1}(C) is a grayscale plot of the
density of states in the junction area, $D(E)$,
in the plane of $|V_0|$ and energy $E$ of electron.
The band edge of the lowest conduction channel
in the leads is at $E/\ef=0.616$.
A virtual bound state of $P$-like state is seen at the Fermi
level at the first zero point of the conductance
($|V_0|/\ef \approx 2.7$). This indicates a resonant
scattering via the virtual bound state there.
However, we do not observe any virtual bound state at the
second zero point.

We observe a perfect spin polarization of $P_z=\pm 1$
in Fig.\ \ref{fig:T1}(B).
However, the absolute value of the conductance is
very small when $P_z=\pm 1$. Hence, it should be difficult
to apply this situation to a spin filter.


\begin{thebibliography}{99}

\bibitem{Zutic}
I.\ \v{Z}uti\'c, J.\ Fabian, and S.\ Das Sarma,
Rev.\ Mod.\ Phys.\ {\bf 76}, 323 (2004).
\bibitem{Winkler}
R.\ Winkler, \textit{Spin-Orbit Coupling Effects in Two-Dimensional
Electron and Hole Systems} (Springer, Berlin Heidelberg, 2003).

\bibitem{Rashba}
E.\ I.\ Rashba, Fiz.\ Tverd.\ Tela (Leningrad) {\bf 2}, 1224
(1960); Yu.\ A.\ Bychkov and E.\ I.\ Rashba, J.\ Phys.\ C {\bf 17},
6039 (1984).
%E.\ I.\ Rashba, Sov.\ Phys.\ Solid State \textbf{2} 1109 (1960).
\bibitem{Nitta}
J.\ Nitta, T.\ Akazaki, H.\ Takayanagi, and T.\ Enoki,
Phys.\ Rev.\ Lett.\ {\bf 78}, 1335 (1997).
\bibitem{Grundler}
D.\ Grundler, Phys.\ Rev.\ Lett.\ {\bf 84}, 6074 (2000).
\bibitem{Yamada}
Y.\ Sato, T.\ Kita, S.\ Gozu, and S.\ Yamada,
J.\ Appl.\ Phys.\ {\bf 89}, 8017 (2001).
\bibitem{spintransistor}
S.\ Datta and B.\ Das, Appl.\ Phys.\ Lett.\ \textbf{56}, 665 (1990).
\bibitem{mismatch}
G.\ Schmidt, D.\ Ferrand, L.\ W.\ Molenkamp, A.\ T.\ Filip,
and B.\ J.\ van Wees, Phys.\ Rev.\ B \textbf{62}, R4790 (2000).

\bibitem{Bulgakov}
E.\ N.\ Bulgakov, K.\ N.\ Pichugin, A.\ F.\ Sadreev,
P.\ Streda, and P.\ Seba,
Phys.\ Rev.\ Lett.\ \textbf{83}, 376 (1999).
\bibitem{Kiselev}
A.\ A.\ Kiselev and K.\ W.\ Kim, Appl.\ Phys.\ Lett.\ \textbf{78},
775 (2001).
\bibitem{Kiselev2}
A.\ A.\ Kiselev and K.\ W.\ Kim, J.\ Appl.\ Phys.\ \textbf{94},
4001 (2003).
\bibitem{Pareek}
T.\ P.\ Pareek, Phys.\ Rev.\ Lett.\ \textbf{92}, 076601 (2004).
 
 
\bibitem{3diode}
T.\ Koga, J.\ Nitta, H.\ Takayanagi, and S.\ Datta,
Phys.\ Rev.\ Lett.\ \textbf{88}, 126601 (2002).
\bibitem{Eto05}
M.\ Eto, T.\ Hayashi, and Y.\ Kurotani,
J.\ Phys.\ Soc.\ Jpn.\ \textbf{74}, 1934 (2005).
\bibitem{Silvestrov}
P.\ G.\ Silvestrov and E.\ G.\ Mishchenko,
Phys.\ Rev.\ B {\bf 74}, 165301 (2006).
\bibitem{3termSG}
J.\ I.\ Ohe, M.\ Yamamoto, T.\ Ohtsuki, and J.\ Nitta,
Phys.\ Rev.\ B \textbf{72}, 041308(R) (2005).
\bibitem{Krich}
J.\ J.\ Krich and B.\ I.\ Halperin,
Phys.\ Rev.\ B \textbf{78}, 035338 (2008).
\bibitem{Yamamoto}
M.\ Yamamoto and B.\ Kramer, J.\ Appl.\ Phys.\ \textbf{103},
123703 (2008).

\bibitem{Murakami}
S.\ Murakami, N.\ Nagaosa, and S.-C.\ Zhang,
Science \textbf{301}, 1348 (2003).
\bibitem{Sinova}
J.\ Sinova, D.\ Culcer, Q.\ Niu, N.\ A.\ Sinitsyn,
T.\ Jungwirth, and A.\ H.\ MacDonald,
Phys.\ Rev.\ Lett.\ \textbf{92}, 126603 (2004).
\bibitem{Wunderlich}
J.\ Wunderlich, B.\ Kaestner, J.\ Sinova, and T.\ Jungwirth,
Phys.\ Rev.\ Lett.\ \textbf{94}, 047204 (2005).

%------------------
\bibitem{Dyakonov}
M.\ I.\ Dyakonov and V.\ I.\ Perel, Phys.\ Lett.\ \textbf{35A}, 459 (1971).
\bibitem{Hirsch}
J.\ E.\ Hirsch, Phys.\ Rev.\ Lett.\ \textbf{83}, 1834 (1999).
\bibitem{Zhang}
S.\ Zhang, Phys.\ Rev.\ Lett.\ \textbf{85}, 393 (2000).
\bibitem{Engel}
H.\ A.\ Engel, B.\ I.\ Halperin, and E.\ I.\ Rashba,
Phys.\ Rev.\ Lett.\ \textbf{95}, 166605 (2005).

\bibitem{Mott}
N.\ F.\ Mott and H.\ S.\ Massey,
{\it Theory of Atomic Collisions}, 3rd ed.\ (Oxford, 1965).
\bibitem{Landau}
L.\ D.\ Landau and E.\ M.\ Lifshitz,
{\it Quantum Mechanics} (Course of theoretical physics, Vol.\ 3)
3rd ed.\ (Pergamon Press, 1977).

\bibitem{Kato}
Y.\ K.\ Kato, R.\ C.\ Myers, A.\ C.\ Gossard, and D.\ D.\ Awschalom,
Science \textbf{306}, 1910 (2004).
\bibitem{Eto}
M.\ Eto and T.\ Yokoyama,
J.\ Phys.\ Soc.\ Jpn.\ \textbf{78}, 073710 (2009).


\bibitem{Hattori}
K.\ Hattori and H.\ Okamoto, Phys.\ Rev.\ B \textbf{74}, 155321 (2006).
\bibitem{Bellucci}
S.\ Bellucci and P.\ Onorato, Phys.\ Rev.\ B \textbf{74}, 245314 (2006).
\bibitem{Jiang}
Y.\ Jiang and L.\ Hu, Phys.\ Rev.\ B \textbf{74}, 075302 (2006).
\bibitem{Xing}
Y.\ Xing, Q.\ F.\ Sun, L.\ Tang, and J.\ P.\ Hu, Phys.\ Rev.\ B \textbf{74}, 155313 (2006).

\bibitem{Dresselhaus}
G.\ Dresselhaus, Phys.\ Rev.\ \textbf{100}, 580 (1955).


\bibitem{Kouwenhoven}
L.\ P.\ Kouwenhoven, C.\ M.\ Marcus, P.\ L.\ McEuen, S.\
Tarucha, R.\ M.\ Westervelt, and N.\ S.\ Wingreen, in
{\it Mesoscopic Electron Transport}, NATO ASI Series E {\bf 345},
eds.\ L.\ Y.\ Sohn, L.\ P.\ Kouwenhoven, and G.\ Sch\"{o}n
(Kluwer, Dordrechit, 1997), p.\ 105.
\bibitem{Aono}
T.\ Aono, Phys.\ Rev.\ Lett.\ {\bf 93}, 116601 (2004),
and references cited therein for the spin pumping.
\bibitem{Feinberg}
D.\ Feinberg and P.\ Simon,
Appl.\ Phys.\ Lett.\ {\bf 85}, 1846 (2004).
\bibitem{Pustilnik1}
M.\ Pustilnik and L.\ Borda,
Phys.\ Rev.\ B {\bf 73}, 201301(R) (2006).

\bibitem{Sharma}
P.\ Sharma and C.\ Chamon,
Phys.\ Rev.\ Lett.\ \textbf{87}, 096401 (2001).
\bibitem{Citro}
R.\ Citro, N.\ Andrei, and Q.\ Niu,
Phys.\ Rev.\ B \textbf{68}, 165312 (2003).
\bibitem{Pustilnik2}
M.\ Pustilnik, E.\ G.\ Mishchenko, and O.\ A.\ Starykh,
Phys.\ Rev.\ Lett.\ \textbf{97}, 246803 (2006).
\bibitem{Braunecker}
B.\ Braunecker, D.\ E.\ Feldman, and F.\ Li,
Phys.\ Rev.\ B \textbf{76}, 085119 (2007).
\bibitem{Abanin}
Related to the quantum wires, we also refer a paper on a 
spin filter using the spin edge states of quantum Hall effect in graphene;
D.\ A.\ Abanin, P.\ A.\ Lee, and L.\ S.\ Levitov,
Phys.\ Rev.\ Lett.\ \textbf{96}, 176803 (2006).


\bibitem{Datta}
S.\ Dtta, \textit{Electronic Transport in Mesoscopic Systems}
(Cambridge University Press, Cambridge, 1995).
\bibitem{Ando}
T.\ Ando, Phys.\ Rev.\ B \textbf{40}, 5325 (1989).
\bibitem{Ando2}
T.\ Ando, Phys.\ Rev.\ B \textbf{44}, 8017 (1991).
\bibitem{Yamamoto2}
M.\ Yamamoto, T.\ Ohtsuki, and B.\ Kramer, Phys.\ Rev.\ B
\textbf{72}, 115321 (2005).

\bibitem{com1}
For the calculation of DOS, a small imaginary part, $i\ef /100$,
is added to $E$ in Eq.\ (\ref{eq:Green}), to broaden
the sharp peaks corresponding to the bound states.
\bibitem{com2}
This dispersion relation is obtained from Eq.\ (\ref{eq:cosBand})
if the cosine-band is approximated by the parabolic band
(wide limit of the leads).


\bibitem{Fert}
A.\ Fert and O.\ Jaoul, Phys.\ Rev.\ Lett.\ \textbf{28}, 303 (1972).
\bibitem{Fert2}
A.\ Fert, A.\ Friederich, and A.\ Hamzic,
J.\ Magn.\ Magn.\ Mater.\ \textbf{24}, 231 (1981).
\bibitem{Guo}
G.\ Y.\ Guo, S.\ Maekawa, and N.\ Nagaosa,
Phys.\ Rev.\ Lett.\ \textbf{102}, 036401 (2009),
and related references cited therein.

\bibitem{Aharonov-Bohm}
Y.\ Aharonov and D.\ Bohm, Phys.\ Rev.\ \textbf{115}, 485 (1959).

\end{thebibliography}
\end{document}